\documentclass[twocolumn]{aastex62}
\usepackage{amsmath}
\graphicspath{{./}{figures/}}

\shorttitle{Are There Terrestrial Planets Lurking in the Outer Solar System?}
\shortauthors{A. Siraj}

\begin{document}

\title{Are There Terrestrial Planets Lurking in the Outer Solar System?}

\email{siraj@princeton.edu}

\author{Amir Siraj}
\affil{Department of Astrophysical Sciences, Princeton University, 4 Ivy Lane, Princeton, NJ 08540, USA}




\begin{abstract}

Motivated by recent measurements of the free-floating planet mass function at terrestrial masses, we consider the possibility that the solar system may have captured a terrestrial planet early in its history. We show that $\sim 1.2$ captured free-floating planets with mass strictly greater than that of Mars may exist in the outer solar system, with a median predicted distance of $\sim 1400 \mathrm{\; AU}$. If we consider a logarithmic bin centered on the mass of Mars, rather than a cutoff, we find that $\sim 2.7$ captured free-floating planets with mass comparable to Mars may exist in the outer solar system. We derive an expectation value of $\sim 0.9$ for the number of captured free-floating planets with mass comparable to that of Mars ($\sim 1.4$ for mass comparable to that of Mercury) that are currently brighter than the 10-year co-added point source detection limits of the Vera C. Rubin Observatory’s Legacy Survey of Space and Time (LSST). Blind shift-and-stack searches could potentially enable the detection of such a planet if it is currently in the Southern sky. The theoretical argument presented here does not rely on the existence of posited patterns in the orbital elements of small bodies in and beyond the Kuiper belt, in contrast with other hypothetical outer-solar-system planets motivated in recent years.

\end{abstract}



\section{Introduction}

The first free-floating planets were discovered over two decades ago \citep{2000MNRAS.314..858L}. Since then, microlensing searches have become the favored method for free-floating planet detection \citep{2011Natur.473..349S, 2017Natur.548..183M, 2019ApJS..244...29M}. In recent years, the first low-mass free-floating planets have been detected \citep{2020ApJ...903L..11M, 2022JKAS...55..173G}. Most recently, \cite{2023AJ....166..108S} measured the mass function of free-floating planets down to low-mass planets. Here, we use this mass distribution to ask the question: could there exist yet-undiscovered terrestrial planets in the outer solar system that were gravitationally captured by the solar system early in its history?

The possibility explored here is entirely distinct from Planet Nine and from other hypothetical planets \citep{2008AJ....135.1161L, 2016AJ....151...22B, 2016MNRAS.455L.114W, 2017AJ....154...62V, 2018AJ....155...75S, 2023AJ....166..118L}. Planet Nine, a proposed $\sim 6 \mathrm{\; M_{\oplus}}$ planet with a semi-major axis of $\sim 400 \mathrm{\; AU}$, is a possibility motivated by observed clustering of extreme trans-Neptunian objects in the outer solar system \citep{2016AJ....151...22B, 2016ApJ...824L..23B, 2019PhR...805....1B, 2021AJ....162..219B}. The potential for its existence is debated \citep{2017AJ....154...50S, 2020ApJS..247...32B, 2020PSJ.....1...28B, 2020AJ....159..285C, 2020AJ....160...50Z, 2021PSJ.....2...59N}. Several searches for Planet Nine have been and continue to be conducted, including ones with the Wide-field Infrared Survey Explorer \citep{2018AJ....155..166M}, the Transiting Exoplanet Survey Satellite \citep{2020PSJ.....1...81R}, the Atacama Cosmology Telescope \citep{2021ApJ...923..224N}, the Zwicky Transient Facility \citep{2022AJ....163..102B}, and the Dark Energy Survey \citep{2022AJ....163..216B}. Here, we present theoretical arguments for yet-undiscovered sub-Earth-mass planets, motivated by novel statistics on free-floating planets. This contrasts with the Planet Nine hypothesis' basis in the posited clustering of solar system objects, and Planet Nine's mass several times that of the Earth. 

In this \textit{Letter}, we consider the implications of recently discovered free-floating planets for the possibility that the solar system hosts terrestrial planets at large distances from the Sun. In Section \ref{theory}, we create a theoretical framework for exploring this possibility and explore the masses of such planets. In Section \ref{detectability}, we evaluate the detectability of such planets. Finally, in Section \ref{discussion}, we discuss the results in the context of the literature and highlight key areas for future work.

\section{Theory}
\label{theory}

\cite{2017MNRAS.472L..75P} studied the capture of free-floating planets by the solar system in its birth cluster, and considered three scenarios for the birth environment: subvirial (partially collapsing), slightly supervirial (gently expanding), or highly supervirial (unbound, rapidly expanding). \cite{2019PhR...805....1B} noted that planet capture is enhanced for supervirial clusters but also that solar system enrichment of short-lived radiogenic isotopes is suppressed for those clusters (and enhanced for subvirial ones, in which planet capture is suppressed).\footnote{We note that the abundance of $^{26}\mathrm{Al}$ in the \cite{2017MNRAS.472L..75P} simulations could potentially increase if the sumlation timespan was increased from 10 Myr to 100 Myr}. While planetary capture and expulsion rates should balance each other out for a shorter simulation (leading to an accurate estimate of retained planets), enrichment from supernovae would appear suppressed. To be as conservative as possible, we analyze the subvirial case, in which planet capture is suppressed but observed meteoric enrichment is more easily explained.

The numerical approach adopted in \cite{2017MNRAS.472L..75P} improves on the earlier calculations of planetary capture \citep{2016ApJ...823L...3L, 2016MNRAS.460L.109M} in part by including Gaussian kinematic substructure, as opposed to assuming a Maxwellian distribution. We adopt the recently quantified mass function of free-floating planets \citep{2023AJ....166..108S} to estimate the mass of a captured free-floating planet with an expected abundance of unity in the solar system as a function of semi-major axis,

\begin{equation}
M_p = 8 M_{\oplus} \left[2.18^{+0.52}_{-1.40} f_{ffp} f(<a) \right]^{1 / \alpha} \; \; ,
\label{ffpeq}
\end{equation}
where the fraction of free-floating planets that are captured is $f_{ffp} \sim 2\%$,\footnote{See Figure 1 of \cite{2017MNRAS.472L..75P}. Given that the capture fraction for 10 $\mathrm{M_{\oplus}}$ planets is higher than the fraction for Jupiter-mass planets, we expect that the capture fraction for planets considered here ($\sim$Mercury-mass) may be above $2\%$, but we conservatively adopt $f_{ffp} = 2\%$ here.} and where the cumulative fraction of captured planets out to a certain semi-major axis $a$, $f(<a)$, is adopted from Figure 2 of \cite{2017MNRAS.472L..75P}. We adopt the power-law slope of $\alpha = 0.96^{+0.47}_{-0.27}$, which was applied by \cite{2023AJ....166..108S} down to Mars-mass planets. In Equation \eqref{ffpeq} and throughout this \textit{Letter}, unless otherwise stated, we adopt $d\mathrm{N} / d\log{\mathrm{M}}$, meaning that the quoted abundance of planets corresponds to planets masses within a logarithmic bin centered on the quoted mass.

Survival in the field is essentially guaranteed given survival in the birth cluster since the low stellar density and high velocity dispersion vastly outweigh the longer timescale \citep{2019PhR...805....1B}. 

\begin{figure}
 \centering
\includegraphics[width=1\linewidth]{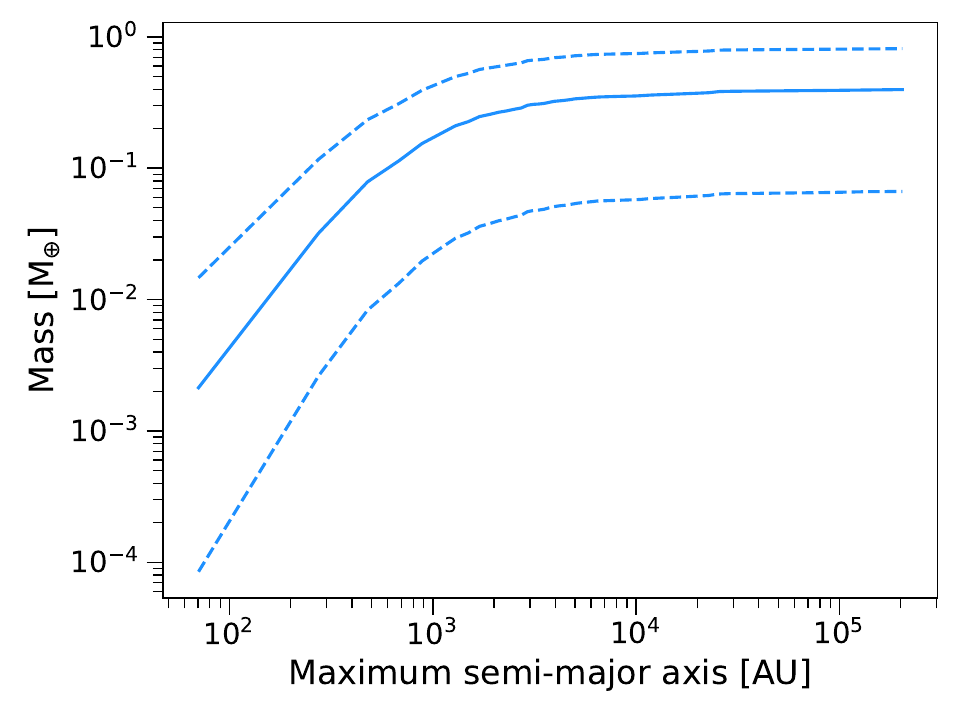}
\caption{Mass of a captured free-floating planet with an expected abundance of unity in the solar system as a function of maximum semi-major axis. Dashed lines correspond to the $25^{\mathrm{th}}$ and $75^{\mathrm{th}}$ percentile limits.}
\label{fig:captured_planets}
\end{figure}

To evaluate the expected mass of the most massive captured free-floating planet as a function of maximum semi-major axis, given the arguments presented in Section \ref{theory}, we implement a Monte Carlo simulation. In order to compute $M_p$ for a given choice of $a$, as described in Equation \eqref{ffpeq}, we draw the normalization factor interior to the brackets from a split Gaussian distribution centered at 2.18 with $\sigma_u = 0.52$ and $\sigma_l = 1.40$ and we draw $\alpha$ from a split Gaussian distribution centered at 0.96 with $\sigma_u = 0.47$ and $\sigma_l = 0.27$. As mentioned in Section \ref{theory}, we adopt $f_{ffp} = 2\%$ and the $f(<a)$ distribution for the subvirial case in \cite{2017MNRAS.472L..75P}. We evaluate $M_p$ $10^6$ times out to each value of $a$ (which is sampled $10^3$ times in log-space) to obtain their probability distributions. The median and $50 \%$ confidence interval are displayed in Figure \ref{fig:captured_planets}. 

\section{Detectability}
\label{detectability}

\begin{figure}
 \centering
\includegraphics[width=1\linewidth]{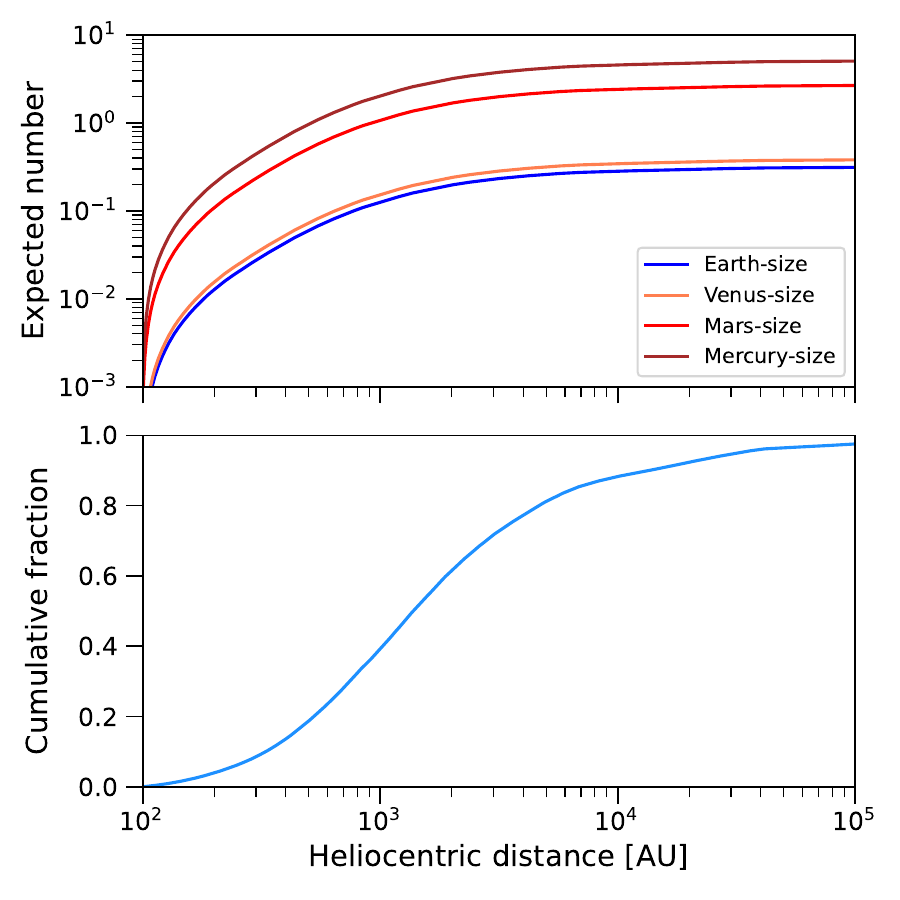}
\caption{Lower panel: cumulative fraction of capture free-floating planets as a function of heliocentric distance. This is independent of planet size. Upper panel: Expected number of Earth-, Venus-, Mars-, and Mercury-size planets as a function of heliocentric distance, adopting a nominal albedo of 0.2.}
\label{fig:heliocentric}
\end{figure}

\begin{figure}
 \centering
\includegraphics[width=1\linewidth]{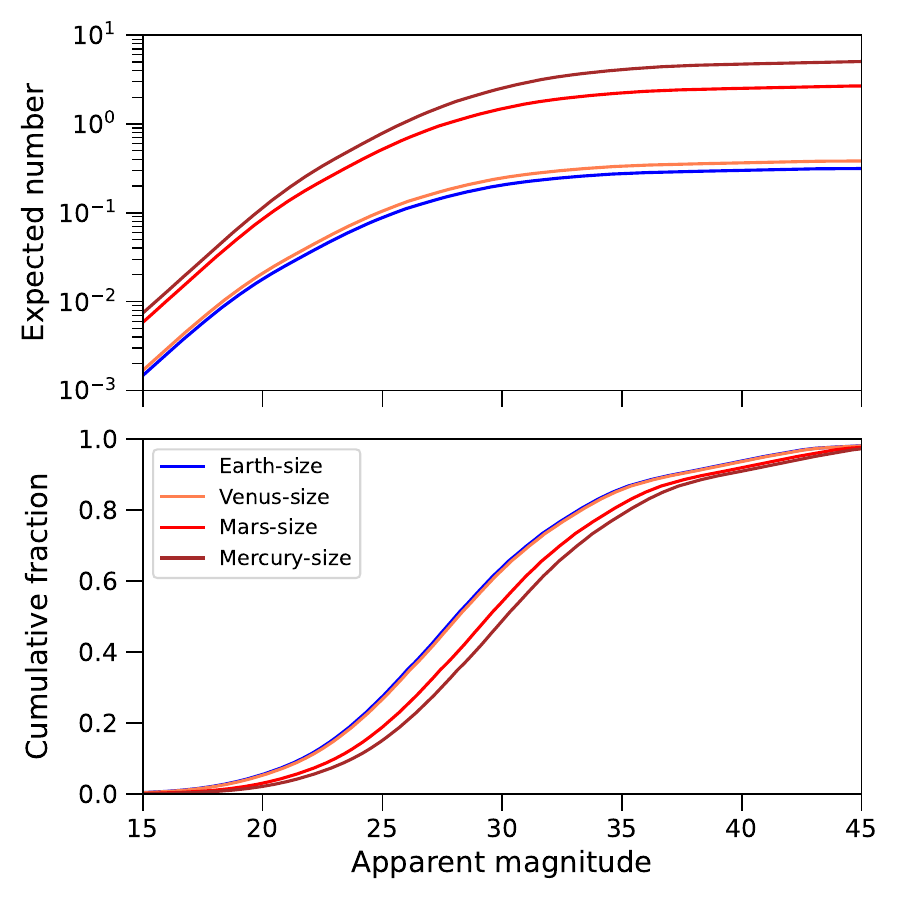}
\caption{Lower panel: cumulative fraction of capture free-floating planets as a function of apparent visual magnitude for Earth-, Venus-, Mars-, and Mercury-size planets. Upper panel: Expected number of planets as a function of apparent visual magnitude, adopting a nominal albedo of 0.2.}
\label{fig:apparent}
\end{figure}

We use a Monte Carlo simulation to explore the detectability of captured planets in the outer solar system. Firstly, we draw instances of semi-major axis $a$ and eccentricity $e$ for the probability distributions in \cite{2017MNRAS.472L..75P} corresponding to the subvirial case. Next, we draw a value for the planet's mean anomaly $M$ from a uniform distribution between 0 and $2 \pi$ (implying that the observation occurs at an arbitrary time). We then calculate the planet's eccentric anomaly by solving Kepler's equation $M = E - e \sin{E}$, and subsequently compute the planet's heliocentric distance, $a(1 - e \cos{E})$. We run this routine $10^8$ times and record the heliocentric distance for each run. Figure \ref{fig:heliocentric} shows the cumulative probability distribution of heliocentric distances (lower panel), as well as this distribution multiplied by the capture fraction, $f_{ffp} \sim 2\%$, and by the best-fit power-law for $d\mathrm{N} / d\log{\mathrm{M}}$ from \cite{2023AJ....166..108S} evaluated at the masses of the four terrestrial planets, to give the expected number of planets within logarithmic mass bins centered on the quoted mass (upper panel). To directly evaluate detectability, we calculate the distribution of apparent magnitudes using the distribution of heliocentric distances, the geometric cross-sections of the four terrestrial planets, and a nominal albedo of $0.2$. Figure \ref{fig:apparent} shows the cumulative probability distribution of apparent magnitudes for each planet size (lower panel), as well as the expected number of planets (upper panel), calculated the same way as in Figure \ref{fig:heliocentric}. Note that the upper panels of both Figures \ref{fig:heliocentric} and \ref{fig:apparent} adopt the best-fit values in \cite{2023AJ....166..108S}, which correspond to the solid blue curve in Figure \ref{fig:captured_planets}.

We estimate that the number of captured free-floating planets in the outer solar system with mass strictly greater than that of Mars is $\sim 1.2$, and that the number of such planets with a strict cutoff at the mass of Mercury is $\sim 2.4$.\footnote{These numbers are derived by evaluating the integral of the expression in Equation 15 of \cite{2023AJ....166..108S} with a lower bound corresponding to the limiting mass (Mercury or Mars mass) and an upper bound of 20 Jupiter masses, and multiplying by the aforementioned capture fraction of $f_{ffp} \sim 2\%$.} When we instead adopt logarithmic bins centered at the Mars mass and the Mercury mass, respectively, we find that the expected number of such planets is $\sim 2.7$ for mass comparable to that of Mars and $\sim 5.2$ for mass comparable to that of Mercury. These planets would have a median heliocentric distance of $\sim 1400 \mathrm{\; AU}$, with $\sim$half of them existing in the range $600 - 3500 \mathrm{\; AU}$.

LSST's 10-year co-added detection limits in the $g$ and $r$ bands are ${26.9}$.\footnote{\url{https://www.lsst.org/scientists/keynumbers}} We note that the magnitude limit for detecting a planet in the outer solar system would be lower than this value, in reality, to compensate for inevitable false positives encountered during the blind shift-and-stack search procedure. Simply for the purposes of illustrating the effect of such a penalty, we will adopt a somewhat arbitrary value of $1$ mag.

At the 10-year co-added detection limit of LSST, $26.9$ mag, the expectation values are $\sim 1.4$ Mercury-size planets and $\sim 0.9$ Mars-size planets. If we include a fiducial blind shift-and-stack penalty of $1$ mag (bringing the limit to $25.9$ mag), the expectation values decrease respectively to $\sim 1.0$ Mercury-size planets and $\sim 0.7$ Mars-size planets. Note that these numbers would change if the albedo of these planets were taken to be a value other than the nominal one adopted here, $0.2$. Any captured planets discovered by LSST would likely exist at a heliocentric distance in the range of $400 - 700 \mathrm{\; AU}$.

LSST will only survey about half of the celestial sphere, so if the closest captured planet were located in the other half of the sky, it would not be detected. Future work should explore the potential for telescopes in the Northern hemisphere to contribute to the search for the closest captured planet. Additionally, in contrast to Planet Nine, we do not expect the parameter space for a Mercury- or Mars-mass planet in the outer solar system to overlap with the detection capabilities of the Atacama Cosmology Telescope \citep{2021ApJ...923..224N}.

\section{Discussion}
\label{discussion}

\cite{2012ApJ...750...83P} numerically estimated that $3 - 6 \times (N_{ffp} / 1) \%$ of all stars capture a planetary companion at a distance of $\sim \mathrm{few} \times 100 - 10^6 \mathrm{\; AU}$, where $N_{ffp}$ is the number of free-floating per star in the birth clusters. Using the \cite{2023AJ....166..108S} result down to $M_{min} = 0.4 \mathrm{M_{\oplus}}$, this implies $\sim 1 - 2$ planets with mass $\gtrsim 0.4 M_{\oplus}$ captured per star. We note that this estimate is in excellent agreement with our calculations, which suggest that there should be $\sim 1$ planet with mass $\gtrsim 0.4 M_{\oplus}$ in the Oort cloud.

In recent years, proposed outer-solar-system planets have primarily been motivated by posited patterns in the orbital elements of small bodies in and beyond the Kuiper belt \citep{2008AJ....135.1161L, 2016AJ....151...22B, 2016MNRAS.455L.114W, 2017AJ....154...62V, 2023AJ....166..118L}. In contrast, this \textit{Letter} presented a theoretical argument motivated by new observational constraints on the abundance of free-floating planets as a function of mass \citep{2023AJ....166..108S}. While other papers have explored the implications of free-floating planets for the possibility of bound, undiscovered planets in the outer solar system, to date they have produced very low likelihoods for the existence of such planets \citep{2016ApJ...823L...3L, 2016MNRAS.460L.109M, 2017MNRAS.472L..75P}. We showed, based on a straightforward theoretical argument, that captured terrestrial planets are likely to exist in the outer solar system. This prediction is unique in terms of the combination of semi-major axes and masses of the proposed planets. Additionally, the prediction avoids the numerous pitfalls involved in identifying patterns in the orbital elements of Kuiper belt objects (e.g. \citealt{2017AJ....154...50S, 2019AJ....158...49V, 2020ApJS..247...32B, 2020PSJ.....1...28B, 2021PSJ.....2...59N}).

Future work should include simulations studying in greater detail the capture and retention of free-floating planets as well as planets bound to other stars. In addition, simulations can shed light on the probability distribution for the orbital plane and position on the sky for captured planets. Future work should also explore other observational tests for the existence of captured planets. If the closest captured planet is currently in the Southern sky and has favorable orbital conditions for detection, it could potentially be discovered in LSST images with a sufficiently advanced blind shift-and-stack algorithm. The detectability of captured planets with surveys like LSST should be explored in more detail. We note that in the future, discoveries of free-floating planets with the Nancy Grace Roman Telescope \citep{2020AJ....160..123J} and detections of extreme trans-Neptunian objects with LSST \citep{2004AAS...20510822I} will help refine the estimates made in this paper. If the statistics from \cite{2023AJ....166..108S} hold to significantly lower masses, it may even be possible to detect captured dwarf planets (with masses comparable to the mass of Pluto) early in the LSST observing program.

\section*{Acknowledgements}
We thank the anonymous referee for insightful comments that greatly improved the quality of the manuscript.





\bibliography{bib}{}
\bibliographystyle{aasjournal}



\end{document}